\documentclass[aps,secnumarabic,nobalancelastpage,
nofootinbib,onecolumn,12pt]{revtex4}%
\usepackage{amsfonts}
\usepackage{amsmath}
\usepackage{graphics}
\usepackage{graphicx}
\usepackage{longtable}
\usepackage{url}
\usepackage{bm}
\usepackage{amssymb}%
\begin{document}
\title{A GPU Poisson-Fermi Solver for Ion Channel Simulations}
\author{Jen-Hao Chen}
\affiliation{Institute of Computational and Modeling Science, National Tsing Hua
University, Hsinchu 300, Taiwan. E-mail: jhchen@mail.nd.nthu.edu.tw}
\author{Ren-Chuen Chen}
\affiliation{Department of Mathematics, National Kaohsiung Normal University, Kaohsiung
802, Taiwan. E-mail: rcchen@nknucc.nknu.edu.tw }
\author{Jinn-Liang Liu}
\affiliation{Institute of Computational and Modeling Science, National Tsing Hua
University, Hsinchu 300, Taiwan. E-mail: jinnliu@mail.nd.nthu.edu.tw}
\date{\today }

\begin{abstract}

\end{abstract}
\maketitle

\textbf{Abstract.} The Poisson-Fermi model is an extension of the classical
Poisson-Boltzmann model to include the steric and correlation effects of ions
and water treated as nonuniform spheres in aqueous solutions.
Poisson-Boltzmann electrostatic calculations are essential but computationally
very demanding for molecular dynamics or continuum simulations of complex
systems in molecular biophysics and electrochemistry. The graphic processing
unit (GPU) with enormous arithmetic capability and streaming memory bandwidth
is now a powerful engine for scientific as well as industrial computing. We
propose two parallel GPU algorithms, one for linear solver and the other for
nonlinear solver, for solving the Poisson-Fermi equation approximated by the
standard finite difference method in 3D to study biological ion channels with
crystallized structures from the Protein Data Bank, for example. Numerical
methods for both linear and nonlinear solvers in the parallel algorithms are
given in detail to illustrate the salient features of the CUDA (compute
unified device architecture) software platform of GPU in implementation. It is
shown that the parallel algorithms on GPU over the sequential algorithms on
CPU (central processing unit) can achieve 22.8$\times$ and 16.9$\times$
speedups for the linear solver time and total runtime, respectively.


\section{Introduction}

Poisson-Boltzmann (PB) solvers are computational kernels of continuum and
molecular dynamics simulations on electrostatic interactions of ions, atoms,
and water in biological and chemical systems
\cite{SH90,HD96,BS01,DN04,CC05,BB09}. The state-of-the-art graphics processing
unit (GPU) with enormous arithmetic capability and streaming memory bandwidth
is now a powerful engine for scientific as well as industrial computing
\cite{OH08,ND10}. In various applications ranging from molecular dynamics,
fluid dynamics, astrophysics, bioinformatics, to computer vision, a CPU
(central processing unit) plus GPU with CUDA (compute unified device
architecture) can achieve 10-137$\times$ speedups over CPU alone \cite{ND10}.

The Poisson-Fermi (PF) model
\cite{L13,LE13,LE14,LE14a,LE15,LE15a,LH16,XL16,LX17} is a fourth-order
nonlinear partial differential equation (PDE) that, in addition to the
electric effect, can be used to describe the steric, correlation, and
polarization effects of water molecules and ions in aqueous solutions. Ions
and water are treated as hard spheres with different sizes, different
valences, and interstitial voids, which yield Fermi-like distributions that
are bounded above for any arbitrary (or even infinite) electric potential at
any location of the system domain of interest. These effects and properties
cannot be described by the classical Poisson-Boltzmann theory that
consequently has been slowly modified and improved
\cite{WR84,DM90,HN95,A95,A96,V99,NO00,K07,GT08,BK09,E11} for more than 100
years since the work of Gouy and Chapman \cite{G10,C13}. It is shown in
\cite{LE14,LX17} that the PF model consistently reduces to the nonlinear PB
model (a second-order PDE) when the steric and correlation parameters in the
PF model vanish. The PF model has been verified with experimental, molecular
dynamics, or Monte Carlo results on various chemical or biological examples in
the above series of papers.

However, in addition to the computational complexity of PB solvers for
biophysical simulations, the PF model incurs more difficulties in numerical
stability and convergence and is thus computationally more expensive than the
PB model as described and illustrated in \cite{L13,LE15}. To reduce long
execution times of PF solver on CPU, we propose here two GPU algorithms, one
for linear algebraic system solver and the other for nonlinear PDE solver. The
GPU linear solver implements the biconjugate gradient stabilized method
(BiCGSTAB) \cite{V92} with Jacobi preconditioning \cite{BB94} and is shown to
achieve 22.8$\times$ speedup over CPU for the linear solver time in a PF
simulation of a sodium/calcium exchanger \cite{LL12}, which is a membrane
protein that removes calcium from cells using the gradient of sodium
concentrations across the cell membrane. The GPU nonlinear solver plus the
linear solver can achieve 16.9$\times$ speedup over CPU for the total
execution time, which shows an improvement of 7$\sim$10$\times$ speedups in
previous GPU studies for Poisson, linear PB, and nonlinear PB solvers
\cite{SK13,CO13}.

The prominent features of CUDA are the thread parallelism on GPU
multiprocessors and the fine-grained data parallelism in shared memory. We use
the standard 3D finite difference method to discretize the PF model with a
simplified matched interface and boundary scheme for the interface condition
\cite{L13}. This structured method allows us to exploit these features as
illustrated in our GPU algorithms.

The rest of this paper describes our algorithms and implementations in more
detail. Section 2 briefly describes the Poisson-Fermi theory and its
application to the sodium/calcium exchanger as an example. Section 3 outlines
all numerical methods proposed in our previous work for the PF model that are
relevant to the GPU algorithms and implementations given in Section 4. Section
5 summarizes our numerical results in comparison of CPU and GPU computing
performances. Concluding remarks are given in Section 6.

\section{Poisson-Fermi Theory}

For an aqueous electrolyte in a solvent domain $\Omega_{s}$ with $K$ species
of ions and water (denoted by $K+1$), the entropy model proposed in
\cite{LE14,LX17} treats all ions and water molecules of any diameter as
nonuniform hard spheres with interstitial voids. Under external field
conditions, the distribution (concentration) of particles in $\Omega_{s}$ is
of Fermi-like type%
\begin{equation}
C_{i}(\mathbf{r})=C_{i}^{\text{B}}\exp\left(  -\beta_{i}\phi(\mathbf{r}%
)+\frac{v_{i}}{v_{0}}S^{\text{trc}}(\mathbf{r})\right)  \text{, \ \ }%
S^{\text{trc}}(\mathbf{r})=\ln\left(  \frac{\Gamma(\mathbf{r)}}{\Gamma
^{\text{B}}}\right)  , \tag{1}%
\end{equation}
since it saturates \cite{LE14}, i.e., $C_{i}(\mathbf{r})<\frac{1}{v_{i}}$ for
any arbitrary (or even infinite) electric potential $\phi(\mathbf{r})$ at any
location $\mathbf{r\in}$ $\Omega_{s}$ for all $i=1,$ $\cdots,$ $K+1 $ (ions
and water), where $\beta_{i}=q_{i}/k_{B}T$ with $q_{i}$ being the charge on
species $i$ particles and $q_{K+1}=0$, $k_{B}$ is the Boltzmann constant, $T$
is an absolute temperature, $v_{i}=4\pi a_{i}^{3}/3$ with radius $a_{i}$, and
$v_{0}=\left(  \sum_{i=1}^{K+1}v_{i}\right)  /(K+1)$ an average volume. The
steric potential $S^{\text{trc}}(\mathbf{r})$ \cite{L13} is an entropic
measure of crowding or emptiness at $\mathbf{r}$ with $\Gamma(\mathbf{r)}%
=1-\sum_{i=1}^{K+1}v_{i}C_{i}(\mathbf{r})$ being a function of void volume
fractions and $\Gamma^{\text{B}}=1-\sum_{i=1}^{K+1}v_{i}C_{i}^{\text{B}}$ a
constant bulk volume fraction of voids when $\phi(\mathbf{r})=0 $ that yields
$\Gamma(\mathbf{r)}=\Gamma^{\text{B}}$, where $C_{i}^{\text{B}}$ are constant
bulk concentrations. The factor $v_{i}/v_{0}$ in Eq. (1) shows that the steric
energy $\frac{-v_{i}}{v_{0}}S^{\text{trc}}(\mathbf{r})k_{B}T$ of a type $i$
particle at $\mathbf{r} $ depends not only on the steric potential
$S^{\text{trc}}(\mathbf{r})$ but also on its volume $v_{i}$ similar to the
electric energy $\beta_{i}\phi(\mathbf{r})k_{B}T$ depending on both the
electric potential $\phi(\mathbf{r})$ and its charge $q_{i}$ \cite{LX17}. The
steric potential is a mean-field approximation of Lennard-Jones (L-J)
potentials that describe local variations of L-J distances (and thus empty
voids) between every pair of particles. L-J potentials are highly oscillatory
and extremely expensive and unstable to compute numerically.

A nonlocal electrostatic formulation of ions and water is proposed in
\cite{LX17} to describe the correlation effect of ions and the polarization
effect of polar water. The formulation yields the following fourth-order
Poisson equation \cite{S06}%
\begin{equation}
\epsilon_{s}\epsilon_{0}l_{c}^{2}\Delta(\Delta\phi(\mathbf{r}))-\epsilon
_{s}\epsilon_{0}\Delta\phi(\mathbf{r})=\rho(\mathbf{r}),\text{\ }\mathbf{r}%
\in\Omega_{s}\text{,} \tag{2}%
\end{equation}
that accounts for electrostatic, correlation, polarization, nonlocal, and
excluded volume effects in electrolytes with only one parameter $l_{c}$ called
correlation length, where $\Delta=\nabla\cdot\nabla$ is the Laplace operator,
$\epsilon_{0}$ is the vacuum permittivity, $\epsilon_{s}$ is a dielectric
constant in the solvent domain, and $\rho(\mathbf{r})=\sum_{i=1}^{K}q_{i}%
C_{i}(\mathbf{r})$ is ionic charge density.

The sodium/calcium exchanger (NCX) structure in an outward-facing conformation
crystallized by Liao et al. \cite{LL12} from \textit{Methanococcus jannaschii}
(NCX\_Mj with the PDB \cite{BW00} code 3v5u) is shown in Fig. 1 and used as an
example for our discussions in what follows. All numerical methods proposed
here are not restricted to this example and can be applied to more general
model problems in electrolyte solutions, ion channels, or other biomolecules.
The NCX consists of 10 transmembrane (TM) helices in which 8 helices (TMs 2 to
5 and 7 to 10 labeled numerically in the figure) form in the center a tightly
packed core that consists of four cation binding sites (binding pocket)
arranged in a diamond shape and shown by three green (putative Na$^{+}$
binding sites) and one blue (Ca$^{2+}$ site) spheres \cite{LH16}. The crystal
structure with a total of $M=4591$ charged atoms is embedded in the protein
domain $\Omega_{p}$ of 10 TMs while the binding sites are in the solvent
domain $\Omega_{s}$. Fig. 2 illustrates a cross section of 3D simulation
domain $\overline{\Omega}=\overline{\Omega}_{s}\cup\overline{\Omega}_{m}$,
where the solvent domain $\overline{\Omega}_{s}$ consists of extracellular and
intracellular baths (in white) and Na$^{+}$ (green) and Ca$^{2+}$ (blue)
pathways, and the biomolecular domain $\overline{\Omega}_{m}$ (yellow)
consists of cell membrane (without charges) and NCX protein ($\Omega_{p}$)
\cite{LH16}. \begin{figure}[t]
\centering\includegraphics[scale=0.7]{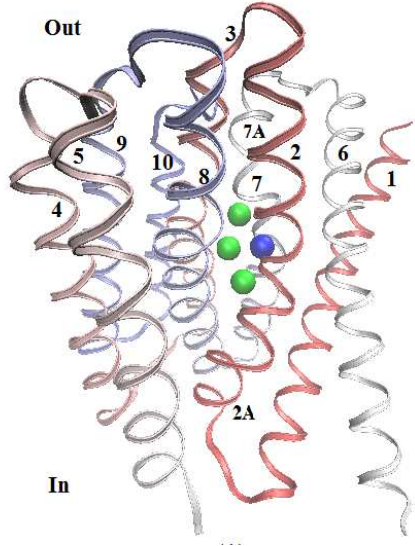}\caption{Structure of NCX\_Mj
viewed from the membrane. The four binding sites shown as four spheres are
tightly formed by 8 transmembrane helices labeled by 2 to 5 and 7 to 10. The
three green and one blue spheres illustrate three putative Na$^{+}$ binding
sites and one Ca$^{2+}$ site, respectively \cite{LH16}.}%
\end{figure}

The electric potential $\phi(\mathbf{r})$ from the structure in the
biomolecular domain $\Omega_{m}$ is described by the Poisson equation%
\begin{equation}
-\epsilon_{m}\epsilon_{0}\nabla^{2}\phi(\mathbf{r})=\sum_{j=1}^{M}q_{j}%
\delta(\mathbf{r}-\mathbf{r}_{j})\text{, }\forall\mathbf{r}\in\Omega
_{m}\text{,}\tag{3}%
\end{equation}
where $\epsilon_{m}$ is the dielectric constant of biomolecules, $q_{j}$ is
the charge of the $j^{\text{th}}$ atom in the NCX protein obtained by the
software PDB2PQR \cite{DN04}, and $\delta(\mathbf{r}-\mathbf{r}_{j})$ is the
Dirac delta function at $\mathbf{r}_{j}$, the coordinate \cite{LL12} of that
atom. The boundary and interface conditions for $\phi(\mathbf{r})$ in $\Omega$
are%
\begin{equation}
\left\{
\begin{array}
[c]{l}%
\phi(\mathbf{r})=0\text{ on }\partial\Omega_{D}=\left\{  \mathbf{r}\in
\partial\Omega\text{: }z=-7\text{ or }z=53\text{ \r{A}}\right\}  ,\\
\nabla\phi(\mathbf{r})\cdot\mathbf{n}=0\text{ on }\partial\Omega_{N}=\left\{
\mathbf{r}\in\partial\Omega\text{: }x=\pm20\text{ or }y=\pm20\text{ \r{A}%
}\right\}  ,\\
\left[  \phi(\mathbf{r})\right]  =\left[  \epsilon\epsilon_{0}\nabla
\phi(\mathbf{r})\cdot\mathbf{n}\right]  =0\text{ on }\partial\Omega_{m}%
\cap\partial\Omega_{s},
\end{array}
\right.  \tag{4}%
\end{equation}
where $\mathbf{n}$ is an outward normal unit vector, $\left[  u(\mathbf{r}%
)\right]  $ is a jump function across the interface $\partial\Omega_{m}%
\cap\partial\Omega_{s}$ between $\Omega_{s}$ and $\Omega_{m}$, $\epsilon
=\epsilon_{s}$ in $\Omega_{s}$, and $\epsilon=\epsilon_{m}$ in $\Omega_{m}$
\cite{L13}.\begin{figure}[t]
\centering\includegraphics[scale=0.7]{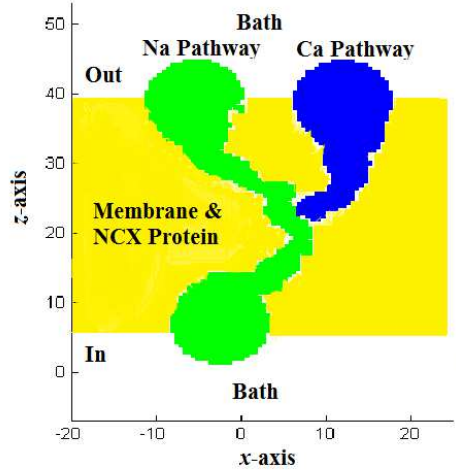}\caption{A cross section of 3D
simulation domain $\overline{\Omega}=\overline{\Omega}_{s}\cup\overline
{\Omega}_{m}$, where the solvent domain $\overline{\Omega}_{s}$ consists of
extracellular and intracellular baths (in white) and Na$^{+}$ (green) and
Ca$^{2+}$ (blue) pathways, and the biomolecular domain $\overline{\Omega}_{m}$
(yellow) consists of cell membrane and NCX protein \cite{LH16}.}%
\end{figure}

\section{Numerical Methods}

To avoid large errors in approximation caused by the delta function
$\delta(\mathbf{r}-\mathbf{r}_{j})$ in Eq. (3), the potential function can be
decomposed as \cite{L13,CL03,GY07}%
\begin{equation}
\phi(\mathbf{r})=\left\{
\begin{array}
[c]{l}%
\widetilde{\phi}(\mathbf{r})+\phi^{\ast}(\mathbf{r})+\phi^{\text{L}%
}(\mathbf{r})\text{\ in }\Omega_{m}\\
\widetilde{\phi}(\mathbf{r})\text{ in }\Omega_{s},
\end{array}
\right.  \tag{5}%
\end{equation}
where $\phi^{\ast}(\mathbf{r})=\sum_{j=1}^{M}q_{j}/(4\pi\epsilon_{m}\left\vert
\mathbf{r-r}_{j}\right\vert )$ and $\widetilde{\phi}(\mathbf{r})$ is found by
solving%
\begin{align}
\left(  l_{c}^{2}\Delta-1\right)  \nabla\cdot\epsilon_{s}\nabla\widetilde
{\phi}(\mathbf{r})  &  =\rho(\mathbf{r})\text{ in }\Omega_{s}\tag{6}\\
-\nabla\cdot\epsilon_{m}\nabla\widetilde{\phi}(\mathbf{r})  &  =0\text{ in
}\Omega_{m} \tag{7}%
\end{align}
without singular source terms $q_{j}\delta(\mathbf{r}-\mathbf{r}_{j})$ and
with the interface condition%
\begin{equation}
\left[  \epsilon\nabla\widetilde{\phi}(\mathbf{r})\cdot\mathbf{n}\right]
=\epsilon_{m}\nabla\left(  \phi^{\ast}(\mathbf{r})+\phi^{\text{L}}%
(\mathbf{r}))\right)  \cdot\mathbf{n}\text{ on }\partial\Omega_{m}\cap
\partial\Omega_{s}\text{.} \tag{8}%
\end{equation}
The potential function $\phi^{\text{L}}(\mathbf{r})$ is the solution of the
Laplace equation
\begin{equation}
\Delta\phi^{\text{L}}(\mathbf{r})=0\text{ in }\Omega_{m} \tag{9}%
\end{equation}
with the boundary condition%
\begin{equation}
\phi^{\text{L}}(\mathbf{r})=\phi^{\ast}(\mathbf{r})\text{ on }\partial
\Omega_{m}. \tag{10}%
\end{equation}
The evaluation of the Green's function $\phi^{\ast}(\mathbf{r})$ on
$\partial\Omega_{m}$ always yields finite numbers and thus avoids the
singularity in the solution process.

The Poisson-Fermi (PF) equation (6) is a nonlinear fourth-order PDE in
$\Omega_{s}$. Newton's iterative method is usually used for solving nonlinear
problems. We seek the solution $\widetilde{\phi}(\mathbf{r})$ of the
linearized PF equation%
\begin{equation}
\epsilon_{s}\left(  l_{c}^{2}\Delta-1\right)  \Delta\widetilde{\phi}%
-\rho^{\prime}(\widetilde{\phi}_{0})\text{ }\widetilde{\phi}=\rho
(\widetilde{\phi}_{0})-\rho^{\prime}(\widetilde{\phi}_{0})\text{ }%
\widetilde{\phi}_{0}\text{ in }\Omega_{s}\text{,} \tag{11}%
\end{equation}
where $\widetilde{\phi}_{0}(\mathbf{r})$ is given, $\rho(\widetilde{\phi}%
_{0})=\sum_{k=1}^{K}q_{k}C_{k}^{0}(\mathbf{r})$, $C_{k}^{0}(\mathbf{r}%
)=C_{k}^{\text{B}}\exp\left(  -\beta_{k}\widetilde{\phi}_{0}(\mathbf{r}%
)+\frac{v_{k}}{v_{0}}S_{0}^{\text{trc}}(\mathbf{r})\right)  $, $S_{0}%
^{\text{trc}}(\mathbf{r})=\ln\left(  \frac{\Gamma_{0}(\mathbf{r)}}%
{\Gamma^{\text{B}}}\right)  $, $\Gamma_{0}(\mathbf{r)}=1-\sum_{k=1}^{K+1}%
v_{k}C_{k}^{0}(\mathbf{r})$, $\rho^{\prime}(\widetilde{\phi}_{0})=\sum
_{k=1}^{K}\left(  -\beta_{k}q_{k}\right)  C_{k}^{0}(\mathbf{r})$, and
$\rho^{\prime}(\widetilde{\phi})=\frac{d}{d\widetilde{\phi}}\rho
(\widetilde{\phi})$. This linear equation is then solved iteratively by
replacing the old function $\widetilde{\phi}_{0}$ by newly found solution
$\widetilde{\phi}$ and so on until a tolerable approximate potential function
$\widetilde{\phi}$ is reached. Note that the differentiation in $\rho^{\prime
}(\widetilde{\phi}) $ is performed only with respect to $\widetilde{\phi}$
whereas $S^{\text{trc}}$ is treated as another independent variable although
$S^{\text{trc}}$ depends on $\widetilde{\phi}$ as well. Therefore,
$\rho^{\prime}(\widetilde{\phi}_{0})$ is not exact implying that this is an
inexact Newton's method \cite{DE82} that has been shown to be highly efficient
in electrostatic calculations for biological systems \cite{HS95,CH10}.

To avoid numerical complexity in using higher order approximations to the
fourth-order derivative, Eq. (11) is reduced to two second-order PDEs
\cite{L13}%
\begin{align}
\text{PF1}  &  :\text{ \ \ }\epsilon_{s}\left(  l_{c}^{2}\Delta-1\right)
\Psi(\mathbf{r})=\rho(\widetilde{\phi}_{0})\text{ in }\Omega_{s}\tag{12}\\
\text{PF2}  &  :\text{ \ \ }-\epsilon_{s}\Delta\widetilde{\phi}(\mathbf{r}%
)-\rho^{\prime}(\widetilde{\phi}_{0})\text{ }\widetilde{\phi}(\mathbf{r}%
)=-\epsilon_{s}\Psi(\mathbf{r})-\rho^{\prime}(\widetilde{\phi}_{0})\text{
}\widetilde{\phi}_{0}\text{ in }\Omega_{s} \tag{13}%
\end{align}
by introducing a density like variable $\Psi=\Delta\widetilde{\phi}$ for which
the boundary condition is \cite{L13}%
\begin{equation}
\Psi(\mathbf{r})=0\text{ on }\partial\Omega_{s}. \tag{14}%
\end{equation}
Eqs. (7), (12), and (13) are coupled together by Eq. (8) in the entire domain
$\Omega$. These equations are solved iteratively with an initial guess
potential $\widetilde{\phi}_{0}$. Note that the linear PDE (11) (or
equivalently (12) and (13)) converges to the nonlinear PDE (6) if
$\widetilde{\phi}_{0}$ converges to the exact solution $\widetilde{\phi}$ of
Eq. (6).

The standard 7-point finite difference (FD) method is used to discretize all
elliptic PDEs (7), (9), (12), and (13), where the interface condition (8) is
handled by the simplified matched interface and boundary (SMIB) method
proposed in \cite{L13}. For simplicity, the SMIB method is illustrated by the
following 1D linear Poisson equation (in $x$-axis)
\begin{equation}
-\frac{d}{dx}\left[  \epsilon(x)\frac{d}{dx}\widetilde{\phi}(x)\right]
=f(x)\text{ in }\Omega\tag{15}%
\end{equation}
with the interface condition%
\begin{equation}
\left[  \epsilon\widetilde{\phi}^{\prime}\right]  =-\epsilon_{m}\frac{d}%
{dx}\phi^{\ast}(x)\text{ at }x=\xi=\text{ }\partial\Omega_{m}\cap
\partial\Omega_{s}\text{,} \tag{16}%
\end{equation}
where $\Omega=$ $\Omega_{m}\cup\Omega_{s}$, $\Omega_{m}=(0,$ $\xi)$,
$\Omega_{s}=(\xi,$ $L)$,\ $f(x)=0$ in $\Omega_{m}$, $f(x)\neq0$ in $\Omega
_{s}$, and $\widetilde{\phi}^{\prime}=\frac{d}{dx}\widetilde{\phi}(x)$. The
corresponding cases to Eqs. (7), (8), and (13) in $y$- and $z$-axis follow in
a similar way. Let two FD grid points $x_{l}$ and $x_{l+1}$ across the
interface point $\xi$ be such that $x_{l}<\xi<x_{l+1}$ with $\Delta
x=x_{l+1}-x_{l}$, a uniform mesh, for example, as used in this work. The FD
equations of the SMIB method at $x_{l}$ and $x_{l+1}$ are%
\begin{align}
\epsilon_{m}\frac{-\widetilde{\phi}_{l-1}+(2-c_{1})\widetilde{\phi}_{l}%
-c_{2}\widetilde{\phi}_{l+1}}{\Delta x^{2}}  &  =f_{l}+\frac{c_{0}}{\Delta
x^{2}}\tag{17}\\
\epsilon_{s}\frac{-d_{1}\widetilde{\phi}_{l}+(2-d_{2})\widetilde{\phi}%
_{l+1}-\widetilde{\phi}_{l+2}}{\Delta x^{2}}  &  =f_{l+1}+\frac{d_{0}}{\Delta
x^{2}}, \tag{18}%
\end{align}
where%
\begin{align*}
c_{1}  &  =\frac{\epsilon_{m}-\epsilon_{s}}{\epsilon_{m}+\epsilon_{s}}\text{,
}c_{2}=\frac{2\epsilon_{s}}{\epsilon_{m}+\epsilon_{s}}\text{, }c_{0}%
=\frac{-\epsilon_{m}\Delta x\left[  \epsilon\widetilde{\phi}^{\prime}\right]
}{\epsilon_{m}+\epsilon_{s}}\text{, }\\
d_{1}  &  =\frac{2\epsilon_{m}}{\epsilon_{m}+\epsilon_{s}}\text{, }d_{2}%
=\frac{\epsilon_{s}-\epsilon_{m}}{\epsilon_{m}+\epsilon_{s}}\text{, }%
d_{0}=\frac{-\epsilon_{s}\Delta x\left[  \epsilon\widetilde{\phi}^{\prime
}\right]  }{\epsilon_{m}+\epsilon_{s}}\text{,}%
\end{align*}
$\widetilde{\phi}_{l}$ is an approximation of $\widetilde{\phi}(x_{l})$, and
$f_{l}=f(x_{l})$. Note that the jump value $\left[  \epsilon\widetilde{\phi
}^{\prime}\right]  $ at $\xi$ is calculated exactly since the derivative of
$\phi^{\ast}$ is given analytically.

After discretization in 3D, the Laplace equation (9), the first PF equation
(PF1) (12), or the second PF equation (PF2) (13) coupled with the Poisson
equation (7), together with their respective boundary or interface conditions,
results in a linear system of algebraic equations $\mathbf{A}\Phi=\mathbf{b}$,
where $\mathbf{A}$ is an $N\times N$ symmetric (for Eq. (9) and PF1) or
nonsymmetric (for PF2 due to the interface condition) matrix and $\Phi$ and
$\mathbf{b}$ are $N\times1$ unknown and known vectors, respectively. The
matrix size $N$ equals to the total number of PD grid points $\left\{
\mathbf{r}_{1},\cdots,\mathbf{r}_{N}\right\}  $ and depends on the protein
size and the uniform mesh size $\Delta x$\ in all three axes. For the
simulation domain $\Omega$ given in (4), the matrix size is $N=6,246,961$ with
$\Delta x=0.5$ \AA .

For the NCX protein, the radii of the entrances of the four binding sites in
the Na$^{+}$ pathway are about 1.1 \AA \ \cite{LH16}. The distances between
binding ions and the charged oxygens of chelating amino acid residues are in
the range of 2.3 $\sim$ 2.6 \AA \ \cite{LH16}. Furthermore, the total number
and total charge of these oxygens are 12 and -6.36$e$, respectively
\cite{LH16}. These indicate that the exchange mechanism of NCX should be
investigated at atomic scale. In \cite{LH16}, the following atomic model is
proposed for studying NCX%
\begin{align}
\phi_{b}  &  =\frac{1}{4\pi\epsilon_{0}}\left(  \frac{1}{6}\sum_{k=1}%
^{6}\left(  \sum_{j=1}^{M}\frac{q_{j}}{\epsilon_{p}|c_{j}-A_{k}|}%
+\sum_{m=1,\text{ }m\neq b}^{8}\frac{O_{m}q_{m}}{\epsilon_{b}|c_{m}-A_{k}%
|}\right)  +\frac{O_{b}q_{b}}{\epsilon_{b}a_{b}}\right)  ,\tag{19}\\
S_{b}^{\text{trc}}  &  =\ln\frac{1-\sum_{m=1}^{8}v_{m}/V_{\text{sites}}%
}{\Gamma^{\text{B}}}, \tag{20}%
\end{align}
where $b=$ aS1, bS2, bS3, bS4, aS5, aS6, bS7, or aS8 in Fig. 3, $c_{j}$ is the
center of the $j^{\text{th}}$ atom in the NCX protein, $A_{k}$ is one of six
symmetric surface points on the spherical site $b$ with radius being either
$a_{\text{Na}^{+}}$ or $a_{\text{Ca}^{2+}}$, $|c_{j}-A_{k}|$ is the distance
between $c_{j}$ and $A_{k}$, $q_{m}$ denotes the charge of any other site $m=$
aS1,..., aS8 $\neq b$ if $O_{m}\neq0$ (the site being occupied by an ion,
otherwise $O_{m}=0$), $\epsilon_{p}$ is a dielectric constant for the NCX
protein, $\epsilon_{b}$ is a dielectric constant in $b$, and $v_{m}$ is the
volume of the ion at site $m$ if it is occupied, $v_{m}=0$ otherwise.

The sites bS2, bS3, and bS4 are the three Na$^{+}$ binding (green) sites in
Fig. 1 and bS7 is the Ca$^{2+}$ binding (blue) site. The sites aS1 and aS5 are
two access sites in the Na$^{+}$ pathway to the binding sites whereas aS6 and
aS8 are access sites in the Ca$^{2+}$ pathway as shown in Fig. 3. The
coordinates of the binding sites (bS2, bS3, bS4, bS7) and the access sites
(aS1, aS5, aS6, aS8) are determined by the crystallized structure in
\cite{LL12} and the empirical method in \cite{LH16}, respectively. These eight
sites are numbered by assuming exchanging paths in which Na$^{+}$ ions move
inwards from extracellular to intracellular bath and Ca$^{2+}$ ions move
outwards from intracellular to extracellular bath without changing direction.
The bidirectional ion exchange suggests a conformational change between the
outward- (Fig. 3A) and inward-facing (Fig. 3B) states of NCX \cite{LL12}. The
outward-facing structure is shown in Figs. 1 and 2. The inward-facing
structure has not yet been seen in X-ray, but Liao et al. \cite{LL12} have
proposed an intramolecular homology model by swapping TMs 6-7A and TMs 1-2A
(in Fig. 1) helices to create an inwardly facing structure of NCX\_Mj.
Numerically, we simply reverse the z-coordinate of all protein atoms in Fig. 2
with respective to the center point of the NCX structure \cite{LH16}%
.\begin{figure}[t]
\centering\includegraphics[scale=0.6]{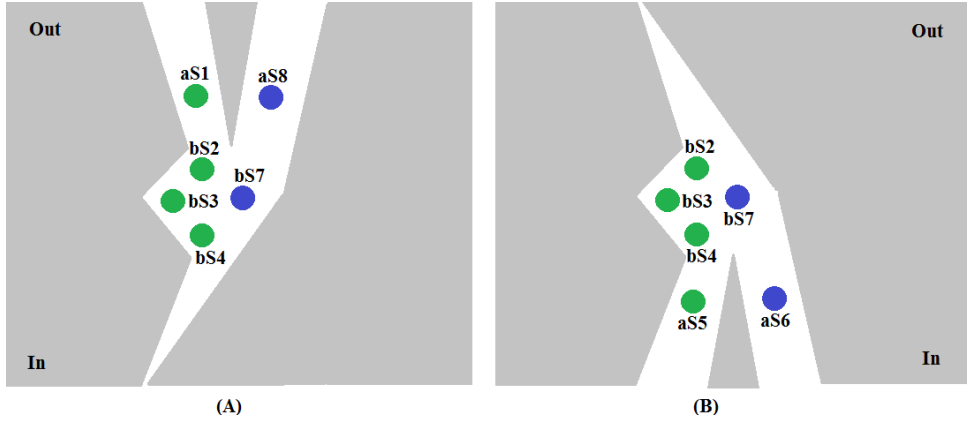}\caption{In (A) outward-facing
structure and (B) inward-facing homology conformations, the eight sites aS1,
bS2,$...$, aS8 marked in green and blue disks located in the Na$^{+}$ (green)
and Ca$^{2+}$ (blue) pathways, respectively, are used for electrostatic
analysis, where bS2, bS3, bS4, and bS7 are binding sites shown in Fig. 1 and
aS1, aS5, aS6, and aS8 are access sites to the binding sites (binding
pocket).}%
\end{figure}

It is shown in \cite{LH16} that numerical results produced by the
Poisson-Fermi model agreed with experimental results on the 3Na$^{+}%
:2$Ca$^{2+}$ stoichiometry of NCX, i.e., NCX extrudes 2 intracellular
Ca$^{2+}$ ions across the cell membrane against [Ca$^{2+}$] gradient in
exchange with 3 extracellular Na$^{+}$ ions using only the energy source of
[Na$^{+}$] downhill gradient. The stoichiometric mechanism described by PF is
based on a transport cycle of state energy changes by the electric and steric
forces on Na$^{+}$ and Ca$^{2+}$ ions occupying or unoccupying their
respective access or binding sites in the NCX structure. The energy state of
each occupied or unoccupied site was obtained by the electric and steric
formulas (19) and (20). Five energy (total potential) states (TPS) have been
proposed and calculated to establish the cyclic exchange mechanism in
\cite{LH16}. The 5 TP states (with their occupied sites) are TPS1 (bS3, bS7),
TPS2 (aS1, bS3, bS7), TPS3 (aS1, bS2, bS4, bS7), TPS4 (bS2, bS3, bS4), and
TPS5 (bS2, bS3, bS4), where TPS1 and TPS5 are in the inward-facing
configuration and TPS2, TPS3 and TPS4 in the outward-facing configuration as
shown in Fig. 4. The transport cycle is a sequence of changing states in the
following order: TPS1 $\rightarrow$ TPS2 $\rightarrow$ TPS3 $\rightarrow$ TPS4
$\rightarrow$ TPS5 $\rightarrow$ TPS1 as shown in Fig. 4. TPS1 is changed to
TPS2 and TPS3 when the extracellular bath concentration $\left[  \text{Na}%
^{+}\right]  _{\text{o}}$ of sodium ions is sufficiently large for one more
Na$^{+}$ to occupy aS1 (in TPS2) or even two more Na$^{+}$ to occupy aS1 and
bS2 (TPS3) such that these two Na$^{+}$ ions have sufficient (positive) energy
to extrude the Ca$^{2+}$ at bS7 in TPS3 out of the binding pocket to become
TPS4. It is postulated in \cite{LH16} that the conformational change of NCX is
induced by the absence or presence of Ca$^{2+}$ at the binding site bS7 as
shown in Fig. 4. After NCX is changed from the outward-facing configuration in
TPS4 to the inward-facing configuration in TPS5, a Ca$^{2+}$ can access aS6
and then move to bS7 in TPS1 if the intracellular bath concentration $\left[
\text{Ca}^{2+}\right]  _{\text{i}}$ of calcium ions is sufficiently large.
Using Eqs. (1), (19), and (20), the selectivity ratio of Na$^{+}$ to Ca$^{2+}$
by NCX from the extracellular bath to the binding site bS2 in TPS1 (after the
conformational change) is defined and given as \cite{LH16}
\begin{equation}
\frac{C_{\text{Na}^{+}}(\mathbf{r})}{C_{\text{Ca}^{2+}}(\mathbf{r})}%
=\frac{\left[  \text{Na}^{+}\right]  _{\text{o}}\exp\left(  -q_{\text{Na}^{+}%
}\Psi_{\text{bS2}}^{\text{TPS1}}\right)  }{\left[  \text{Ca}^{2+}\right]
_{\text{o}}\exp\left(  -q_{\text{Ca}^{2+}}\Psi_{\text{bS2}}^{\text{TPS1}%
}\right)  }=55.4\tag{21}%
\end{equation}
under the experimental bath conditions $\left[  \text{Na}^{+}\right]
_{\text{o}}$ and $\left[  \text{Ca}^{2+}\right]  _{\text{o}}$ given in Table
1, where $\Psi_{b}=(q_{b}\phi_{b}-S_{b}^{\text{trc}}k_{B}T)/q_{b}$. The
selectivity ratio of Ca$^{2+}$ to Na$^{+}$ by NCX from the intracellular bath
to the binding site bS7 in TPS4 is
\begin{equation}
\frac{C_{\text{Ca}^{2+}}(\mathbf{r})}{C_{\text{Na}^{+}}(\mathbf{r})}%
=\frac{\left[  \text{Ca}^{2+}\right]  _{\text{i}}\exp\left(  -q_{\text{Ca}%
^{2+}}\Psi_{\text{bS7}}^{\text{TPS4}}\right)  }{\left[  \text{Na}^{+}\right]
_{\text{i}}\exp\left(  -q_{\text{Na}^{+}}\Psi_{\text{bS7}}^{\text{TPS4}%
}\right)  }=4986.1.\tag{22}%
\end{equation}
\begin{figure}[t]
\centering\includegraphics[scale=1.0]{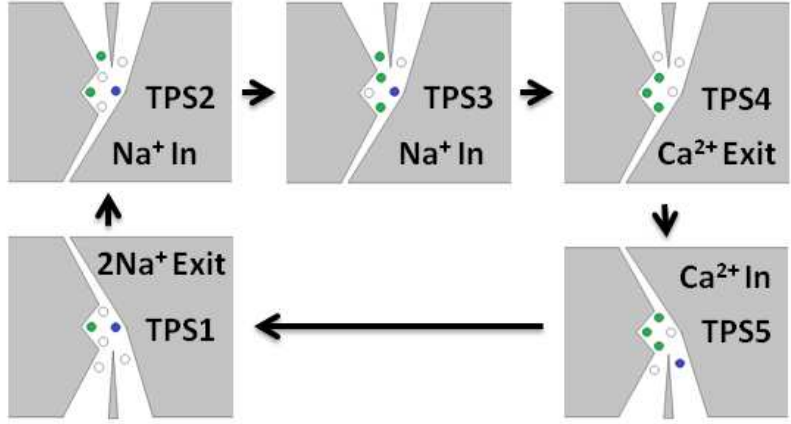}\caption{ A transport cycle of
changing total potential (TP) states of Na ions (green) and Ca ions (blue)
entering and exiting the binding pocket in the order TPS1 $\rightarrow$ TPS2
$\rightarrow$ TPS3 $\rightarrow$ TPS4 $\rightarrow$ TPS5 $\rightarrow$ TPS1 is
proposed in \cite{LH16} to establish the NCX stoichiometric mechanism.}%
\end{figure}

\begin{center}
$%
\begin{tabular}
[c]{c|c|c|c}%
\multicolumn{4}{c}{Table 1: Values of Model Notations}\\\hline
Symbol & Meaning & \ Value & \ Unit\\\hline
\multicolumn{1}{l|}{$k_{B}$} & \multicolumn{1}{|l|}{Boltzmann constant} &
\multicolumn{1}{|l|}{$1.38\times10^{-23}$} & J/K\\
\multicolumn{1}{l|}{$T$} & \multicolumn{1}{|l|}{temperature} &
\multicolumn{1}{|l|}{$298.15$} & K\\
\multicolumn{1}{l|}{$e$} & \multicolumn{1}{|l|}{proton charge} &
\multicolumn{1}{|l|}{$1.602\times10^{-19}$} & C\\
\multicolumn{1}{l|}{$\epsilon_{0}$} & \multicolumn{1}{|l|}{permittivity of
vacuum} & \multicolumn{1}{|l|}{$8.85\times10^{-14}$} & F/cm\\
\multicolumn{1}{l|}{$\epsilon_{b}$, $\epsilon_{p}$, $\epsilon_{s}$,
$\epsilon_{m}$} & \multicolumn{1}{|l|}{dielectric constants} &
\multicolumn{1}{|l|}{$10$, $20$, $78.45$, 2} & \\
\multicolumn{1}{l|}{$l_{c}=2a_{i}$} & \multicolumn{1}{|l|}{correlation length}
& \multicolumn{1}{|l|}{$i=\text{Na}^{+}\text{, Ca}^{2+}$} & \AA \\
\multicolumn{1}{l|}{$a_{\text{Na}^{+}}$, $a_{\text{Ca}^{2+}}$} &
\multicolumn{1}{|l|}{radii} & \multicolumn{1}{|l|}{$0.95$, $0.99$} & \AA \\
\multicolumn{1}{l|}{$a_{\text{Cl}^{-}}$, $a_{\text{H}_{2}\text{O}}$} &
\multicolumn{1}{|l|}{radii} & \multicolumn{1}{|l|}{$1.81$, $1.4$} & \AA \\
\multicolumn{1}{l|}{$O_{m}^{b}$} & \multicolumn{1}{|l|}{site occupancy} &
\multicolumn{1}{|l|}{0 or 1} & \\
\multicolumn{1}{l|}{$\left[  \text{Na}^{+}\right]  _{\text{o}}$, $\left[
\text{Na}^{+}\right]  _{\text{i}}$} & \multicolumn{1}{|l|}{bath
concentrations} & \multicolumn{1}{|l|}{120, 60 \cite{LL12}} & mM\\
\multicolumn{1}{l|}{$\left[  \text{Ca}^{2+}\right]  _{\text{o}}$, $\left[
\text{Ca}^{2+}\right]  _{\text{i}}$} & \multicolumn{1}{|l|}{bath
concentrations} & \multicolumn{1}{|l|}{1, 33 \cite{LL12}} & $\mu$M\\
\multicolumn{1}{l|}{i, o} & \multicolumn{1}{|l|}{intra, extracellular} &
\multicolumn{1}{|l|}{} & \\\hline
\end{tabular}
$
\end{center}

We need to extend the energy profile of each TPS in the filter region
$\Omega_{f}$ $\subset\Omega_{s}$, which contains the access and binding sites
shown in Fig. 3, to the entire simulation domain $\Omega$ with boundary
conditions in which the membrane potential and bath concentrations [Ca$^{2+}$]
and [Na$^{+}$] are given, since the exchange cycle inside the binding pocket
is driven by these far field boundary conditions. Therefore, the total energy
of an ion in the filter region in a particular state is determined by all ions
and water molecules in the system with boundary potentials and is calculated
by the continuum model (12) and (13) in $\overline{\Omega}_{s}\backslash
\Omega_{f}$ and the atomic model (19) and (20) in $\Omega_{f}$. The values of
$\phi_{b}$ and $S_{b}^{\text{trc}}$ of the atomic model can be used as
Dirichlet boundary values for the continuum model. Since all ions and water
are treated as hard spheres with interstitial voids, the algebraic steric
potential $S_{b}^{\text{trc}}$ is consistent with the continuous steric
function $S^{\text{trc}}(\mathbf{r})$. The algebraic electric potential
$\phi_{b}$ is based on Coulomb's law in stead of Poisson's theory. Therefore,
the PF theory is a continuum-molecular theory. We refer to \cite{LH16} for
more details.

\section{GPU Algorithms}

We first describe a nonlinear solver of inexact-Newton type in Table 2 for the
PF model (1) and (2) for sequential coding, where $\left\Vert \Phi\right\Vert
_{\infty}$ is the maximum norm of the $N\times1$ vector $\Phi$, $\omega
_{\text{PF}}=0.3$ is a relaxation parameter, and ErrTol $=10^{-3}$ is an error
tolerance for Newton's iteration. The error tolerance of linear solvers is
$10^{-6}$ for all algebraic systems ($\mathbf{A}^{\text{L}}\Phi^{\text{L}%
}=\mathbf{b}^{\text{L}}$, $\mathbf{A}_{0}\Phi_{0}=\mathbf{b}_{0}$,
$\mathbf{A}^{\text{PF1}}\Psi=\mathbf{b}^{\text{PF1}}$, $\mathbf{A}%
^{\text{PF2}}\Phi=\mathbf{b}^{\text{PF2}}$) in Steps 1 - 4. The vectors
$\Phi^{\text{L}}$, $\Phi_{0}$, $\Psi$, and $\Phi$ are approximate solutions of
$\phi^{\text{L}}(\mathbf{r})$, $\widetilde{\phi}_{0}(\mathbf{r})$,
$\Psi(\mathbf{r})$, and $\widetilde{\phi}(\mathbf{r})$, respectively, at FD
grid points $\mathbf{r}\in\left\{  \mathbf{r}_{1},\cdots,\mathbf{r}%
_{N}\right\}  $. The $N\times N$ matrix $\mathbf{A}$ with entries $a_{ij}$ of
each linear system in Table 2 is compressed to seven $N\times1$ vectors A0[i],
$\cdots$, A6[i] for i = 0, ..., $N$-1, where $a_{ij}$ = A0[i] with $j=$ i,
A1[i] with $j=$ i+1 (the east neighboring point of i), A2[i] with $j=$ i-1
(west), A3[i] with $j=$ i-XPts (south), A4[i] with $j=$ i+XPts (north), A5[i]
with $j=$ i-XPts*YPts (downside), or A6[i] with $j=$ i+XPts*YPts (upside), and
XPts and YPts are total numbers of grid points in the $x$- and $y$-axis,
respectively. Therefore, the matrix is stored in a diagonal format without
offset arrays \cite{BG08}. It has been shown in \cite{BG08} that the diagonal
representation of sparse matrices is memory-bandwidth efficient and has high
computational intensity for the matrix-vector multiplication on CUDA.

\begin{center}%
\begin{tabular}
[c]{ll}%
\multicolumn{2}{c}{Table 2: Sequential Algorithm of PF Nonlinear
Solver}\\\hline
1 & Solve Eqs. (9), (10) in $\overline{\Omega}_{m}$ for $\phi^{\text{L}%
}(\mathbf{r})$ in the discrete form $\mathbf{A}^{\text{L}}\Phi^{\text{L}%
}=\mathbf{b}^{\text{L}}$\\
& by a linear solver (LS).\\
2 & Solve PF2 Eqs. (4), (7), (8), (13) with $\rho^{\prime}=\Psi=0$ in
$\overline{\Omega}\backslash\Omega_{f}$ for $\widetilde{\phi}_{0}(\mathbf{r}%
)$\\
& in $\mathbf{A}_{0}\Phi_{0}=\mathbf{b}_{0}$ by LS.\\
3 & Solve PF1 Eq. (12), (14)$\text{ in }\overline{\Omega}_{s}\backslash
\Omega_{f}$ for $\Psi(\mathbf{r})$ in $\mathbf{A}^{\text{PF1}}\Psi
=\mathbf{b}^{\text{PF1}}$ by LS.\\
4 & Solve PF2 Eqs. (4), (7), (8), (13)$\text{ in }\overline{\Omega}%
_{s}\backslash\Omega_{f}$ for $\widetilde{\phi}(\mathbf{r})$ in $\mathbf{A}%
^{\text{PF2}}\Phi=\mathbf{b}^{\text{PF2}}$ by LS.\\
5 & Assign $\Phi:=\omega_{\text{PF}}\Phi_{0}+(1-\omega_{\text{PF}})\Phi$.\\
& If $\left\Vert \Phi-\Phi_{0}\right\Vert _{\infty}$ $>$ ErrTol, $\Phi
_{0}:=\Phi$ (i.e., $\widetilde{\phi}_{0}:=\widetilde{\phi}$) and go to Step 3;
else stop.\\\hline
\end{tabular}

\end{center}

The conjugate gradient (CG) method is one of the most efficient and widely
used linear solvers for the Poisson-Boltzmann equation in biomolecular
applications \cite{HS95,DM89,NH91,WL10}. Since the PF2 linear system is
nonsymmetric, we use the Bi-CGSTAB method of Van der Vorst \cite{V92}, which
is probably the most popular short recurrence method for large-scale
nonsymmetric linear systems \cite{SS10}.

The parallel platform CUDA created by NVIDIA is an application programming
interface software that gives direct access to the GPU's virtual instruction
set and parallel computational elements and works with programming languages
C, C++, and Fortran \cite{A08,NB08}. Our sequential code is written in C++.
GPU programming is substantially simplified by using CUDA. For example, we
only need to replace CPU\_BiCGSTAB() by GPU\_BiCGSTAB() in the same line of
the function call without changing any other parts of the code, i.e., CUDA
generates a sequential code if CPU\_BiCGSTAB() is called or a parallel code if
GPU\_BiCGSTAB() is called. Of course, the function definition of
GPU\_BiCGSTAB() is different from that of CPU\_BiCGSTAB(). We now describe our
GPU implementation of the Bi-CGSTAB method \cite{BB94} (with pointwise Jacobi
preconditioning) as shown by the algorithm in Table 3, where the symbol C: or
G: indicates that the corresponding statement is executed at the host (CPU) or
the device (GPU).

\begin{center}%
\begin{tabular}
[c]{c}%
Table 3: GPU Algorithm 1 for Bi-Conjugate Gradient Stabilized Method\\\hline
\multicolumn{1}{l}{1 C: Copy $\mathbf{A}$, $\mathbf{b}$ and initial guess
$\mathbf{x}^{(0)}$ from host (C) to device (G).}\\
\multicolumn{1}{l}{2 G: Compute $\mathbf{r}^{(0)}=\mathbf{b}-\mathbf{Ax}%
^{(0)}$ and set $\mathbf{\tilde{r}}=\mathbf{r}^{(0)}.$\hspace*{0.22in}}\\
\multicolumn{1}{l}{3 C: \textbf{for} $i=1,2,...$}\\
\multicolumn{1}{l}{4 G:\hspace*{0.22in} $\rho_{i-1}=\mathbf{\tilde{r}}%
^{T}\mathbf{r}^{(i-1)}$ with CUDA Reduction Method (RM) \cite{H07}}\\
\multicolumn{1}{l}{5 C:\hspace*{0.22in} \textbf{if} $i=1$}\\
\multicolumn{1}{l}{6 G:\hspace*{0.22in} $\ \ \ \mathbf{p}^{(i)}=\mathbf{r}%
^{(i-1)}$}\\
\multicolumn{1}{l}{7 C:\hspace*{0.22in} \textbf{else}}\\
\multicolumn{1}{l}{8 G:\hspace*{0.15in} $\ \ \ \ \ \beta_{i-1}=(\rho
_{i-1}\alpha_{i-1})/(\rho_{i-2}\omega_{i-1})$}\\
\multicolumn{1}{l}{9 G:\hspace*{0.15in} $\ \ \ \ \ \mathbf{p}^{(i)}%
=\mathbf{r}^{(i-1)}+\beta_{i-1}(\mathbf{p}^{(i-1)}-\omega_{i-1}\mathbf{v}%
^{(i-1)})$}\\
\multicolumn{1}{l}{10 C: \ \ \ \ \textbf{end}}\\
\multicolumn{1}{l}{11 G: $\ \ \ \ \mathbf{v}^{(i)}=\mathbf{Ap}^{(i)}$}\\
\multicolumn{1}{l}{12 G: \hspace*{0.15in} $\alpha_{i}=\rho_{i-1}%
/\mathbf{\tilde{r}}^{T}\mathbf{v}^{(i)}$ with RM}\\
\multicolumn{1}{l}{13 G: \hspace*{0.15in} $\mathbf{s}=\mathbf{r}%
^{(i-1)}-\alpha_{i}\mathbf{v}^{(i)}$}\\
\multicolumn{1}{l}{14 G: \hspace*{0.15in} $\mathbf{t}=\mathbf{As}$}\\
\multicolumn{1}{l}{15 G: \hspace*{0.15in} $\omega_{i}=\mathbf{t}^{T}%
\mathbf{s}/\mathbf{t}^{T}\mathbf{t}$ with RM}\\
\multicolumn{1}{l}{16 G: \hspace*{0.15in} $\mathbf{x}^{(i)}=\mathbf{x}%
^{(i-1)}+\alpha_{i}\mathbf{p}^{(i)}+\omega_{i}\mathbf{s}$}\\
\multicolumn{1}{l}{17 G: \hspace*{0.15in} $\mathbf{r}^{(i)}=\mathbf{s}%
-\omega_{i}\mathbf{t}$}\\
\multicolumn{1}{l}{18 G: \ \ \ \ \ Evaluate $\left\Vert \mathbf{r}%
^{(i)}\right\Vert _{\infty}$ with RM and then copy $\left\Vert \mathbf{r}%
^{(i)}\right\Vert _{\infty}$ to host.}\\
\multicolumn{1}{l}{19 C: \hspace*{0.15in} \ If $\left\Vert \mathbf{r}%
^{(i)}\right\Vert _{\infty}<$ ErrTol,}\\
\multicolumn{1}{l}{20 G: \ \ \ \ \ \ \ \ copy $\mathbf{x}^{(i)}$ to host and
stop.}\\
\multicolumn{1}{l}{21 C: \textbf{end}}\\\hline
\end{tabular}

\end{center}

The computational complexity of GPU Algorithm 1 is dominated by (A) the
execution time of (i) the matrix-vector products in Steps 2, 11, and 14 and
(ii) the vector inner products in Steps 4, 12, and 15; and (B) the
synchronization time of threads within a block to share data through shared
memory \cite{NB08}. The CPU and GPU systems used for this work are Intel Xeon
E5-1650 and NVIDIA GeForce GTX TITAN X (with 3072 CUDA cores), respectively.
The total numbers of blocks and threads per block defined in our GPU code were
256 and 1024, respectively, with which the inner product of two vectors of
dimension $N=$ 6,246,961, for example, is processed with $256\times1024=$
262,144 threads of execution. A while loop with a stride of 262,144 is hence
incorporated into the kernel in order to visit all vector elements. Each
thread can access to its private \textit{local} memory, its \textit{shared}
memory, and the same \textit{global} memory of all threads \cite{NB08}. The
fine-grained data parallelism of these vectors is expressed by $256$ blocks in
shared memory. The thread parallelism on GPU multiprocessors (cores) is
transparently scaled and scheduled by CUDA \cite{NB08}. In order to keep all
multiprocessors on the GPU busy, the inner product of these very large vectors
(arrays) is performed using a parallel reduction method proposed in
\cite{H07}. Parallel reduction is a fundamental technique to process very
large arrays in parallel blocks by recursively reducing a portion of the array
within each thread block. The inner-product operation reduces two vectors to a
single scalar.

The matrix $\mathbf{A}^{\text{PF2}}$ and the vectors $\mathbf{b}^{\text{PF1}}
$ and $\mathbf{b}^{\text{PF2}}$ in Table 2 (the sequential nonlinear solver)
need to be updated iteratively according to Eqs. (12) and (13). Iterative
switches between the sequential code on CPU for updating these matrix and
vectors (Step 1 in Table 3) and the parallel code on GPU for solving linear
systems (other Steps in Table 3) drastically reduce the parallel performance
of GPU. It is thus crucial to parallelize the nonlinear solver for which we
propose an algorithm in Table 4, where the matrices $\mathbf{A}^{\text{L}}$
(in Step 2), $\mathbf{A}_{0}$ (Step 4), $\mathbf{A}^{\text{PF1}}$ (Step 5),
and $\mathbf{A}^{\text{PF2}}$ (Step 6) are all stored in A0[i], $\cdots$,
A6[i] and constructed on GPU. The corresponding vectors $\mathbf{b}^{\text{L}%
}$, $\mathbf{b}_{0}$, $\mathbf{b}^{\text{PF1}}$, and $\mathbf{b}^{\text{PF2}}
$ are similarly stored in $\mathbf{b}$ and constructed on GPU. Step 1 in Table
3 is now removed. Note that $\mathbf{A}^{\text{PF2}}$, $\mathbf{b}%
^{\text{PF1}}$ and $\mathbf{b}^{\text{PF2}}$ are updated iteratively on GPU
since $\Phi$ (corresponding to $\widetilde{\phi}(\mathbf{r})$) is updated
iteratively. The vector $\mathbf{b}^{\text{IF}}$ corresponding to the
right-hand side of the interface equation (8) is calculated once in Step 3 on
CPU and repeatedly used in Steps 4 and 6 on GPU.

\begin{center}%
\begin{tabular}
[c]{ll}%
\multicolumn{2}{c}{Table 4: GPU Algorithm 2 for Poisson-Fermi Nonlinear
Solver}\\\hline
1 C: & Allocate the vectors A0[i], $\cdots$, A6[i], $\mathbf{b}$,
$\Phi^{\text{L}}$, $\Psi$, $\Phi_{0}$, $\Phi$ and $\mathbf{b}^{\text{IF}}$ on
the global\\
& memory of the GPU.\\
2 G: & Solve Eqs. (9), (10) in $\overline{\Omega}_{m}$ for $\phi^{\text{L}%
}(\mathbf{r})$ once in $\mathbf{A}^{\text{L}}\Phi^{\text{L}}=\mathbf{b}%
^{\text{L}}$ by GPU Algorithm 1.\\
3 C: & Compute the interface vector $\mathbf{b}^{\text{IF}}$ and copy it from
host to device.\\
4 G: & Solve PF2 Eqs. (4), (7), (8), (13) with $\rho^{\prime}=\Psi=0$ in
$\overline{\Omega}\backslash\Omega_{f}$ for $\widetilde{\phi}_{0}(\mathbf{r}%
)$\\
& in $\mathbf{A}_{0}\Phi_{0}=\mathbf{b}_{0}$ by GPU Algorithm 1.\\
5 G: & Solve PF1 Eqs. (12), (14)$\text{ in }\overline{\Omega}_{s}%
\backslash\Omega_{f}$ for $\Psi(\mathbf{r})$ in $\mathbf{A}^{\text{PF1}}%
\Psi=\mathbf{b}^{\text{PF1}}$\\
& by GPU Algorithm 1.\\
6 G: & Solve PF2 Eqs. (4), (7), (8), (13) in $\overline{\Omega}\backslash
\Omega_{f}$ for $\widetilde{\phi}(\mathbf{r})$ in $\mathbf{A}^{\text{PF2}}%
\Phi=\mathbf{b}^{\text{PF2}}$\\
& by GPU Algorithm 1.\\
7 G: & Assign $\Phi:=\omega_{\text{PF}}\Phi_{0}+(1-\omega_{\text{PF}})\Phi$.
If $\left\Vert \Phi-\Phi_{0}\right\Vert _{\infty}$ $>$ ErrTol, $\Phi_{0}%
:=\Phi$ and\\
& go to Step 5; else stop.\\\hline
\end{tabular}

\end{center}

\section{Results}

We first present some physical results obtained by the PF model.\ For TPS3 in
Fig. 4, the electric potential profiles of $\phi(\mathbf{r})$ along the axes
of Na$^{+}$ and Ca$^{2+}$ pathways (Fig. 2) are shown in Fig. 5 in green and
blue curves, respectively. Each curve was obtained by averaging the values of
$\phi(\mathbf{r})$ at cross sections along the axis of the solvent domain
$\Omega_{s}$ that contains both two baths and a pathway. The potential values
at aS1, bS2, bS3, bS4, bS7, and aS8 were obtained by Eq. (19) whereas the
value at aS9 was obtained by Eqs. (12) and (13). These two curves suggest
opposite flows of Na$^{+}$ and Ca$^{2+}$ ions as illustrated in the figure.
Numerical results presented here are only for TPS3 as those of the other four
states in Fig. 4 follow in the same way of calculation with four times more
computational efforts.\begin{figure}[t]
\centering\includegraphics[scale=0.5]{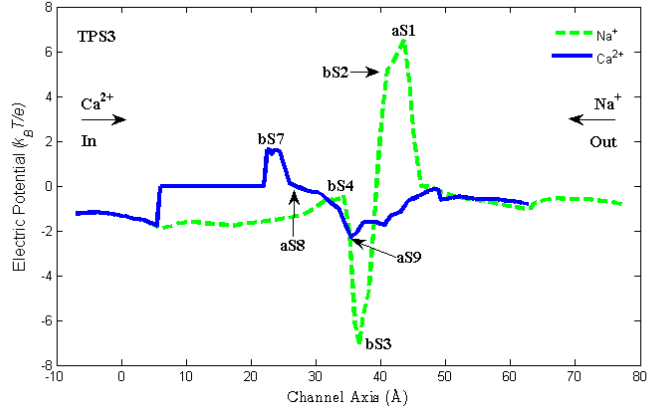}\caption{For TPS3 in Fig. 4,
the electric potential profiles of $\phi(\mathbf{r})$ along the axes of
Na$^{+}$ and Ca$^{2+}$ pathways are shown in green and blue curves,
respectively. Each curve was obtained by averaging the values of
$\phi(\mathbf{r})$ at cross sections along the axis of the solvent domain
$\Omega_{s}$ that contains both two baths and a pathway. The notations aS1,
bS2, bS3, bS4, bS7, aS8, and aS9 indicate the potential values at those sites.
The potential profiles suggest opposite flows of Na$^{+}$ and Ca$^{2+}$ ions
as illustrated in the figure.}%
\end{figure} 

We next show the speedup of the parallel (GPU) computation over the sequential
(CPU) computation in Table 5, where the time is in second, the linear solver
is described in Table 3 for the parallel version that yields the sequential
version by replacing the symbol G: by C:, and the nonlinear solver is
described in Tables 2 and 4 for the CPU and GPU version, respectively. The
speedup in total runtime is drastically reduced from 16.9 to 1890/370 = 5.1 if
the nonlinear solver is not parallelized although the speedup of linear solver
alone is 22.8, where 370 (not shown in the table) is the total runtime of the
mixed algorithm of sequential nonlinear and parallel linear solvers. It is
thus important to parallelize the nonlinear solver of the coupled nonlinear
PDEs, such as the Poisson-Fermi model, in scientific computing since most
realistic applications in engineering or scientific systems are highly
nonlinear. The reduction of speedups from 22.8 (in linear solver time) to 16.9
(in total runtime) is due mainly to the lower efficiency of GPU compared to
that of CPU for constructing matrix systems in Steps 4, 5, and 6 in GPU
Algorithm 2 as shown by the smaller speedup 7.7 (in nonlinear solver time) in
Table 5. Nevertheless, the GPU algorithms of the linear and nonlinear solver
in Tables 3 and 4 for the Poisson-Fermi model improve significantly the
speedups of 7-10 in previous GPU studies for Poisson, linear
Poisson-Boltzmann, and nonlinear Poisson-Boltzmann solvers \cite{SK13,CO13}.

\begin{center}%
\begin{tabular}
[c]{l|l|l|l}%
\multicolumn{4}{c}{Table 5: Speedup of GPU over CPU}\\\hline
& CPU (T1) & GPU (T2) & Speedup (T1/T2)\\\hline
Linear Solver Time in Sec. & \multicolumn{1}{|r|}{$1551$} &
\multicolumn{1}{|r|}{$68$} & \multicolumn{1}{|r}{$22.8$}\\
Nonlinear Solver Time & \multicolumn{1}{|r|}{$339$} &
\multicolumn{1}{|r|}{$44$} & \multicolumn{1}{|r}{$7.7$}\\
Total Runtime & \multicolumn{1}{|r|}{$1890$} & \multicolumn{1}{|r|}{$112$} &
\multicolumn{1}{|r}{$16.9$}\\\hline
\end{tabular}

\end{center}

\section{Conclusion}

We propose two GPU (parallel) algorithms for biological ion channel
simulations using the Poisson-Fermi model that extends the classical
Poisson-Boltzmann model to study not only the continuum but also the atomic
properties of ions and water molecules in highly charged ion channel proteins.
These algorithms exploit the thread and data parallelism on GPU with the CUDA
platform that makes GPU programming easier and GPU computing more efficient.
Numerical methods for both linear and nonlinear solvers in the algorithms are
given in detail to illustrate the salient features of CUDA in implementation.
These parallel algorithms on GPU are shown to achieve 16.9$\times$ speedup
over the sequential algorithms on CPU, which is better than that of previous
GPU algorithms based on the Poisson-Boltzmann model.

\begin{acknowledgments}
This work was supported by the Ministry of Science and Technology, Taiwan
(MOST 106-2115-M-007-010 to J.H.C., 105-2115-M-017-003 to R.C.C., and
105-2115-M-007-016-MY2 to J.L.L.).
\end{acknowledgments}

\end{document}